\documentclass[namedreferences]{solarphysics}
\usepackage[optionalrh,solaromanenum]{spr-sola-addons} 
\usepackage{graphicx}                    
\usepackage{color}                       

\definecolor{NavyBlue}{cmyk}{0.94,0.54,0,0,}
\newcommand{\kms}{km s$^{-1}$}

\chardef\us=`\_

\begin{document}

\begin{article}

\begin{opening}

\title{Temporal and Spatial Scales in Coronal Rain Revealed by UV Imaging and Spectroscopic Observations}

%
\author[addressref={aff1,aff2},corref,email={ryohtaroh.ishikawa@nao.ac.jp}]{\inits{R. T.}\fnm{Ryohtaroh T.}~\lnm{Ishikawa}}
\author[addressref={aff2,aff1}]{\inits{Y. }\fnm{Yukio}~\lnm{Katsukawa}}
\author[addressref=aff3]{\inits{P. }\fnm{Patrick}~\lnm{Antolin}}
\author[addressref=aff4]{\inits{S. }\fnm{Shin}~\lnm{Toriumi}}


%
\runningauthor{R. T. Ishikawa et al.}
\runningtitle{Temporal and Spatial Scales in Coronal Rain}

\address[id={aff1}]{Department of Astronomical Science, School of Physical Sciences, The Gradiate University for Advanced Studies, SOKENDAI, 2-21-1 Osawa, Mitaka, Tokyo 181-8588, Japan}
\address[id={aff2}]{National Astronomical Observatory of Japan, 2-21-1 Osawa, Mitaka, Tokyo 181-8588, Japan}
\address[id={aff3}]{Department of Mathematics, Physics and Electrical Engineering, Northumbria University, Newcastle upon Tyne, NE1 8ST, UK}
\address[id={aff4}]{Institute of Space and Astronautical Science, Japan Aerospace Exploration Agency, 3-1-1 Yoshinodai, Chuo-ku, Sagamihara, Kanagawa 252–5210, Japan}

\begin{abstract}
Coronal rain corresponds to cool and dense clumps in the corona accreting towards the solar surface, 
and is often observed above solar active regions.
They are generally thought to be produced by thermal instability in the corona 
and their lifetime is limited by the time they take to reach the chromosphere.
Although the rain usually fragments into smaller clumps while falling down, 
their specific spatial and temporal scales remain unclear. 
In addition, the observational signatures of the impact of the rain with the chromosphere
have not been clarified yet. 
In this study, we investigate the time evolution of velocity and intensity of coronal rain above a sunspot
by analyzing coronal images obtained by the Atmospheric Imaging Assembly (AIA) onboard the Solar Dynamics Observatory (SDO) 
as well as the Slit-Jaw Images (SJIs) and spectral data taken by the Interface Region Imaging Spectrograph (IRIS) satellite. 
We identify dark and bright threads moving towards the umbra in AIA images and in SJIs, respectively, and co-spatial chromospheric 
intensity enhancements and redshifts in three IRIS spectra, 
Mg {\sc ii} k 2796 \AA, Si {\sc iv} 1394 \AA, and C {\sc ii} 1336 \AA.
The intensity enhancements and coronal rain redshifts occur almost concurrently in all the three lines, 
which clearly demonstrates the causal relationship with coronal rain.
Furthermore, we detect bursty intensity variation with a timescale shorter than 1 minute in Mg {\sc ii} k, Si {\sc iv} and C {\sc ii} spectra, 
indicating that a length scale of rain clumps is about 2.7 Mm
if we multiply the typical time scale of the busty intensity variation at 30 sec by the rain velocity at 90 \kms. 
Such rapid enhancements in the IRIS lines are excited within a time lag of 5.6 sec
limited by the temporal resolution.
These temporal and spatial scales may reflect the physical processes
responsible for the rain morphology, and are suggestive of instabilities such as Kelvin-Helmholtz.

\end{abstract}

%

\end{opening}

%

\section{Introduction}
The solar corona is known to have average temperatures above a million degrees.
However, not all plasmas in the corona have such high temperature: 
plasma at chromospheric temperatures at coronal heights is commonly observed, particularly above active regions.
Coronal rain is one of such cool and dense materials in the corona, usually appearing in chromospheric lines, 
and subsequently falling along a loop-like path into the chromosphere.
Coronal rain was first observed more than 40 years ago \citep{Kawaguchi70, Leroy72}.
Coronal rain also represents the existence of various dynamic, mass and energy transport processes in coronal loops
\citep{Marsch08, McIntosh12}.

Coronal rain is thought to be produced by thermal instability 
during the catastrophic cooling part of a thermal non-equilibrium (TNE) cycle
in which radiative cooling locally overcomes the coronal heating mechanisms
as demonstrated in a one-dimensional numerical simulations 
\citep{Parker1953, Muller03, Muller04, Muller05, Xia11}.
The TNE cycle is induced by concentrated heating localized nearer 
to the chromospheric footpoints than to the loop apex
\citep{Antiochos&Klimchuk91}.
Chromospheric evaporation and density increase in the loop is induced
by the heating, which increases radiative losses in the corona
and results in a loss of thermal equilibrium and catastrophic cooling for sustained footpoint heating.
Recently, \citet{Fang13} and \citet{Fang15} studied 
the formation and the evolution of the coronal rain 
using 2.5-dimensional models and 
showed that multiple clumps in a coronal loop induce siphon flows
which cause shear flows and deform the falling clumps themselves.

The cooling sequence of the loops has been also investigated
using several overvations by the
Transition Region and Coronal Explorer \citep{Schrijver01}
and Extreme Ultraviolet Imaging Telescope onboard 
Solar and Heliospheric Observatory (e.g. \citealt{DeGroof05}),
which detected EUV intensity variations of coronal loops in ARs.
Antolin et al. (2015) showed that chromospheric emission follows the observed cooling in the EUV lines,
as expected from catastrophic cooling events. 
EUV intensity pulsation with long-term
periods were observed in warm coronal loops
\citep{Auchere14,Froment15, Froment17} and shown to match well with TNE cycles.
\citet{Antolin10} argued that coronal rain may point
to the agent of the heating of coronal loops as well as 
the spatial distribution of the heating in loops
using a combination of Hinode observations and loop modelling of Alfv\`enic wave heating.

\citet{Antolin15} coordinated observations with high-resolution and multi-wavelength instruments,
and obtained the specific spatial scale of the coronal rains at each temperature.
The most frequent scales of width and length detected in 1400 {\AA}
slit-jaw images (SJIs) taken by Interface Region Imaging Spectrograph (IRIS: \citealt{DePontieu14})
are $0^{\prime\prime}.88$ and $1^{\prime\prime}.76$, respectively.
\citet{Antolin12b} and \citet{Antolin12a} analyzed off-limb and on-disk $\mathrm{H \alpha}$ observation data 
obtained with the CRisp Imaging SpectroPolarimeter at the Swedish Solar Telescope
and detected cool and dense clump-like structures. 
They interpreted the clumps as coronal rain 
based on their spectral and morphological characteristics. 
In particular they showed that the rain morphology is clumpy and multi-stranded at high resolution, 
with an average length distribution of 700 km with a long tail 
towards longer lengths, and an average width distribution 
of 300 km with a full-width at half maximum (FWHM) of 100 km.
At the same time, they also concluded that there exists undetected, 
even smaller clumps because the current spatial resolution is not enough.

\citet{Kleint14} reported the first observation tracing coronal rain into an umbra 
and associated bright dots with IRIS. 
They interpreted that the coronal rain falls down along loops 
and collides with the local plasma in the transition region (TR), increasing density and temperature in the TR above a sunspot.
In the IRIS spectral lines, rain is observed as bursts of strong Doppler shifts with an average duration of 20s, 
indicating supersonic downward flows of up to 200 \kms just prior to impact,
which is faster than the free fall velocity for a plasma initially at rest.

In this article, we estimate the temporal and spatial scales from observational data obtained by IRIS and SDO.
The rest of the article proceeds as follows. In Section 2, we introduce the data that we use. 
In Section 3, we describe the analysis and results. In Section 4, we summarize and discuss the observational results 
from the view point of scales in the coronal rain.

\section{Observations}

For this study, we analyze the observational data of active region (AR) NOAA 12042 obtained by IRIS and 
the Solar Dynamics Observatory (SDO: \citealt{Pesnell12}).
IRIS observed AR 12042 with the sit-and-stare mode of IRIS and Hinode Operation Plan (IHOP) 250
from 10:00 to 11:15 UT on 24 April 2014,
providing spectral data in three lines (C {\sc ii} 1336 \AA, Si {\sc iv} 1396 \AA, and Mg {\sc ii} k 2796 \AA).
The slit width is $0^{\prime\prime}.33$ and the pixel size and slit length are 
$0^{\prime\prime}.166$ and $119^{\prime\prime}$, respectively. 
IRIS also produced simultaneous SJIs data in 1400 {\AA} and 2796 {\AA},
whose field of view is about $119^{\prime\prime} \times 119^{\prime\prime}$.
The recording cadences of spectral data and SJIs are 5.6 s and 11 s, respectively.
The central sunspot in AR 12042 was located at (x, y)=($475^{\prime\prime}, 359^{\prime\prime}$) at the beginning of the IRIS observation.
We used the level 2 data, where dark-current subtraction, flat-field correction, 
and geometric and wavelength calibration are applied.

The target AR was also covered by the Atmospheric Imaging Assembly (AIA: \citealt{Lemen12}) onboard the SDO satellite, 
which makes synoptic observations in multiple wavelengths in the EUV with a cadence of 12 s. 
In this study, we mainly used the 171 \AA \ channel images to investigate the morphology and evolution.

\section{Results}
\subsection{Overview of the Observed Region}{\label{sec:obverview}}

\begin{figure}[t]
\centerline{\includegraphics[width=13cm]{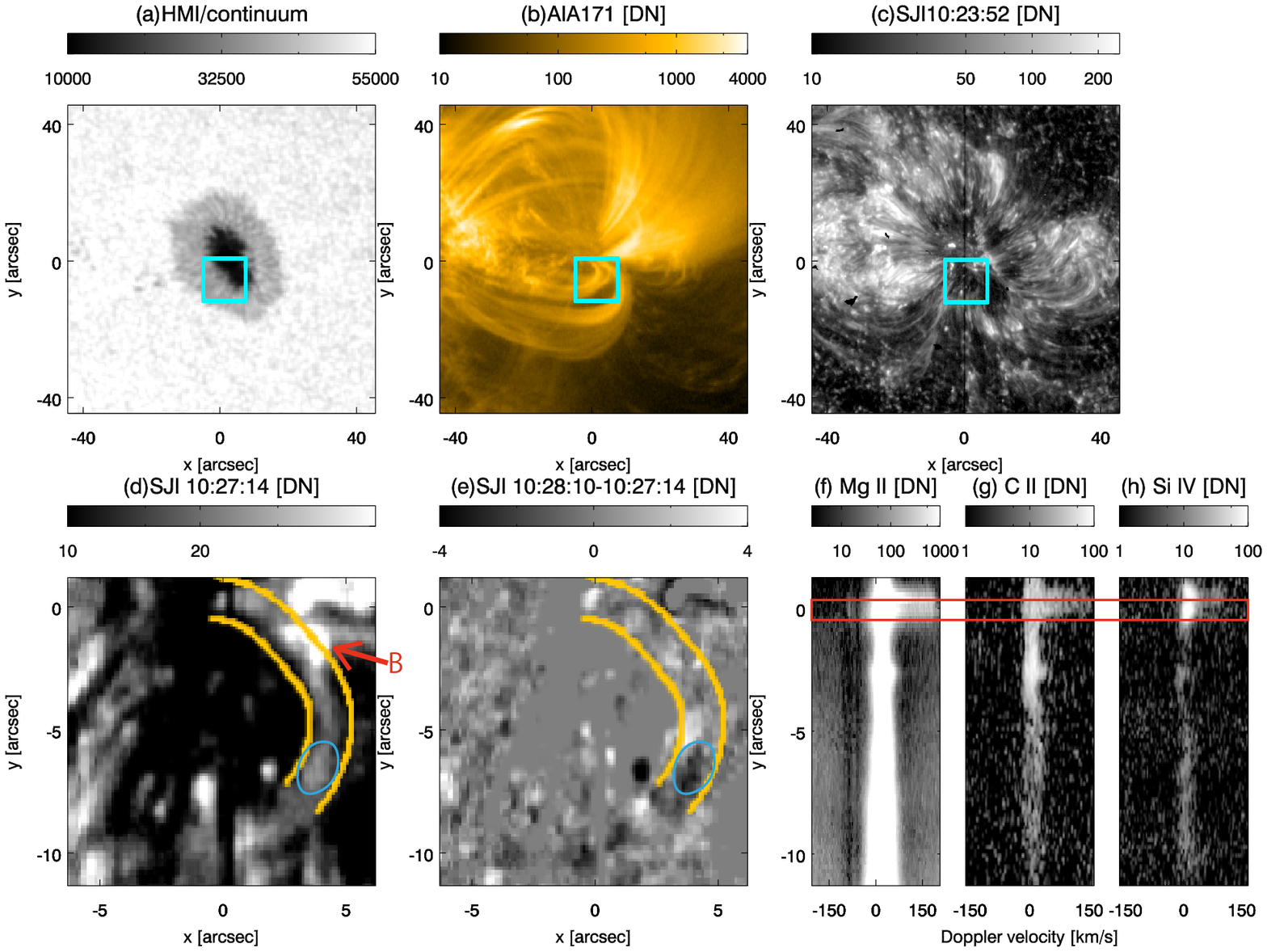}}
\caption{AR NOAA 12042 and coronal rain observed on 24 April 2014.
(a), (b) and (c) show the target AR obtained by SDO/Helioseismic and Magnetic Imager continuum, 
SDO/AIA 171 \AA \ and IRIS/SJI at Si {\sc iv} line (1400 \AA), respectively.
A close-up image of IRIS/SJI at Si {\sc iv} line (panel d)
and a difference between two IRIS/SJI images with a time difference of 56 seconds (panel e)
showing that a bright thread-like structure (indicated by a blue circle)
moves along a loop-like path
(indicated by the yellow curves) to the sunspot umbra.
These yellow curves are determined by the loop-like structure
seen in the AIA 171\AA\ image (Panel b).
The blue rectangles in (a), (b), and (c) correspond to the field-of-view (FOV) of (d) and (e).
The IRIS spectra of Mg {\sc ii} k line 2796 \AA \ (f), Si {\sc iv} line 1396 \AA, and (g) 
and C {\sc ii} line 1336 \AA \ (h) at 10:37:54 UT.
The red rectangle indicates the region where we take the average of the spectra
to obtain the temporal evolution of the intensity of each IRIS spectral line.
The movie corresponding to panels d and f--h can be found in the additional material; this is the file movie1.mpeg.
In this movie we also show the close-up image in AIA 171 {\AA}.}
\label{fig:dataset}
\end{figure}

Panels a, b and c in Figure \ref{fig:dataset} show the image of the active region obtained by
SDO/HMI continuum, 
SDO/AIA at 171 \AA ($\log T=5.8$), and IRIS/SJI at Si {\sc iv} line (1400 \AA, $\log T=4.9$) respectively.
Panel d shows a close-up image of IRIS/SJI at Si {\sc iv} line
and panel e shows a difference between two images of IRIS/SJI,
showing that a bright clump moves along a loop-like trajectory leading into an umbra.
Panels f--h show spectrograph of Mg {\sc ii} k (2796 \AA, $\log T=4$), 
C {\sc ii} (1336 \AA, $\log T=4.4$) and Si {\sc iv} (1396 \AA, $\log T=4.9$)
obtained by IRIS sit-and-stare mode.
The sunspot umbra is located at the center of the FOV shown in panels b--e, and is overlapped by the IRIS slit.
The horizontal axis of the spectral data f--h is the Doppler velocity, 
where positive and negative indicate the red and blueshifts, respectively.

One can see a bright, thread-like structure falling down to the sunspot umbra
along a loop-like path
as indicated by the yellow lines in Figures \ref{fig:dataset}d and e.
The loop-like path coincides with a loop seen in the image in AIA 171 \AA.
This path ends just at the IRIS slit which overlaps the umbra.
In addition, there are some clumps moving along this path (see the movie that accompanies this article, movie1.mpeg).
At the same time, in the IRIS spectral data (panels f--h),
all the three lines show redshifts coinciding with the fall of the clumps
at the footpoint of the trajectory. This indicates that the clumps have
a multi-temperature structure ($4 < \log T < 4.9$, Antolin et al. 2015).
The observed features of this bright thread match the characteristics 
of the coronal rain \citep{Antolin12b, Kleint14}.

\subsection{Temporal Evolution of Coronal Rain}
\label{sec:time_evo}

\begin{figure}[tp]
\centerline{\includegraphics[width=13cm]{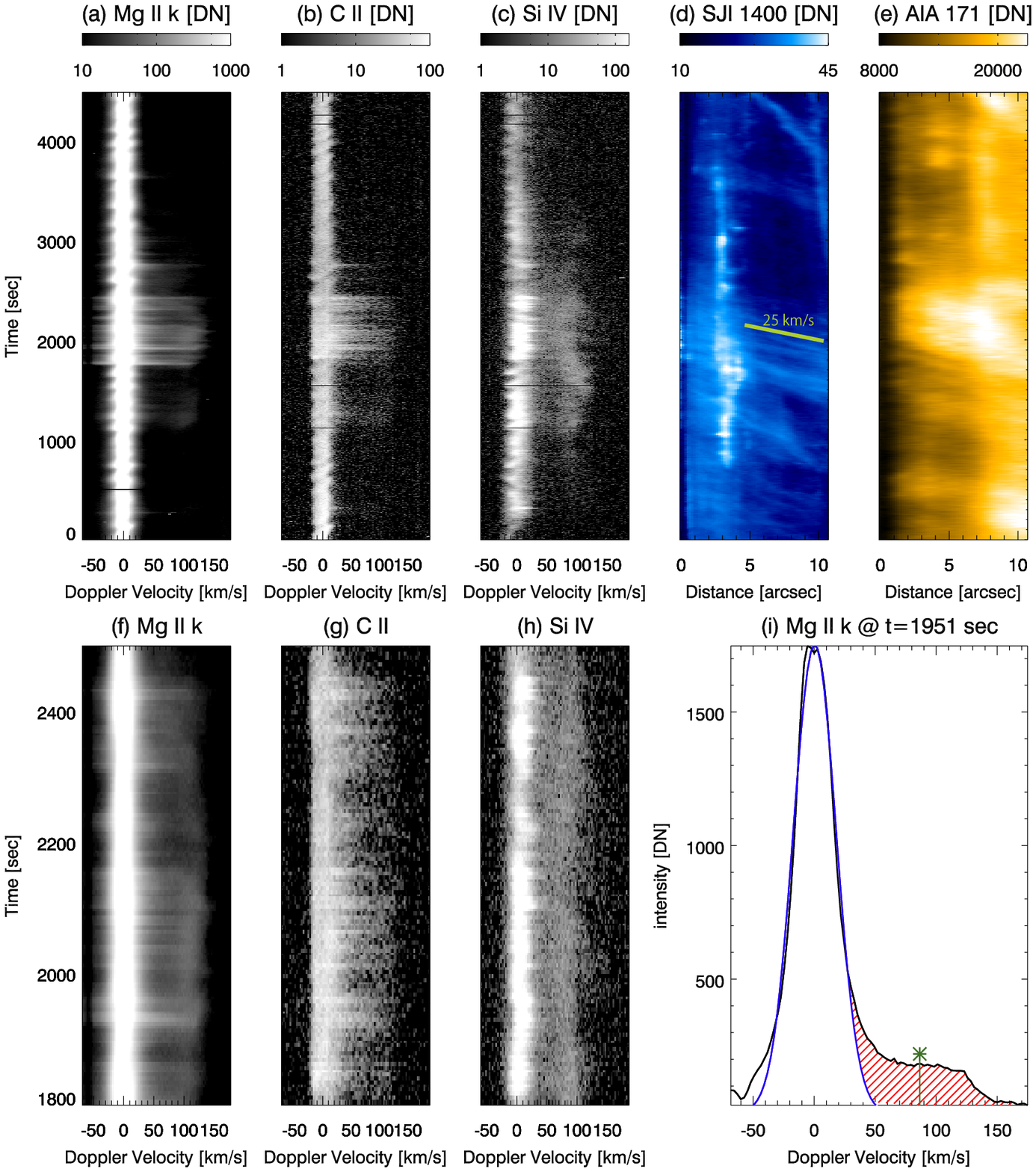}}
\caption{Time evolution for: (a) Mg {\sc ii} k, (b) C {\sc ii}, and (c) Si {\sc iv} lines as a function of Doppler shift.
The origin of time axis is defined as 10:00:28~UT on 24 April 2014.
(d) and (e) show the time evolutions for the IRIS/SJI 1400 \AA \ images and SDO/AIA 171 \AA \ 
along the loop-like trajectory indicated as yellow curves in Figure \ref{fig:dataset}d.
The IRIS slit is located at the origin.
Bright threads (coronal rain) move toward to
the slit and reach there, especially in the range 1800 s $\leq t \leq$ 2500 s.
At the same time, all the lines show large redshifts.
(f), (g), and (h) show close-up time intervals of (a), (b), and (c), respectively.
(i) shows the spectrum of Mg {\sc ii} k line at $t=1951$ s (black line), the stationary component detected by 
a single Gaussian fitting (blue line), and the redshifted component defined as the residual (red region).
The center of gravity is described as a green line in (i), which indicates the redshifted velocity at that moment.}
\label{fig:vt_st}
\end{figure}

Figures \ref{fig:vt_st}a--c show the time evolutions of IRIS spectra
at the footpoint from 10:00:28~UT on 24 April 2014.
We use spectra averaged over 5 pixels along the slit, corresponding to a width of 580 km.
The horizontal axis indicates the Doppler velocity.

Figures \ref{fig:vt_st}d and e show distance-time diagrams along the trajectory of the rain seen in IRIS 1400 \AA \ SJI (Figure \ref{fig:dataset}a) and
AIA 171 \AA, respectively.
We use the distance along the trajectory of the coronal rain indicated by the yellow curves 
in Figures \ref{fig:dataset}d and e,
and define the origin of the rain footpoint on the IRIS slit.
We averaged the data over 11 pixels perpendicular to the trajectory.
In Figure \ref{fig:vt_st}d, we can see some bright clumps, 
indicating the coronal rain, moving toward the umbra along the path in SJI 1400 \AA.
Although there is another bright structure near the slit (at about 2 Mm from the origin; as indicated by B in Figure \ref{fig:dataset}d),
the motion of coronal rain is not affected by the structure, probably because the structure and coronal rain overlap
along the line-of-sight due to the projection effect.

Because there are stationary and highly red-shifted components in the spectra observed with IRIS,
we estimated the Doppler velocity of the coronal rain using the IRIS spectral data with the following method.
First, we defined the central wavelength of the stationary component by applying a Gaussian fit to the IRIS spectra 
and its intensity by integrating the obtained Gaussian profile  (the blue line in Figure \ref{fig:vt_st}i).
After subtracting the stationary component from the original spectra, 
we then measured the integrated intensity and the center of gravity of the residual spectra 
as the intensity and the Doppler velocity of the secondary component representing the coronal rain (the red region in Figure \ref{fig:vt_st}i).

In the IRIS spectra (Figures \ref{fig:vt_st}a--c),
the intensity enhancements and redshift are observed at the same time
when the coronal rain reaches the position of the slit from 1800 to 2500 s, as shown in Figure \ref{fig:vt_st}d.
The mean Doppler velocity of the coronal rain is about 85 \kms in the Si {\sc iv} line, 
while the plane-of-sky velocity along the trajectory (yellow curve in Figures \ref{fig:dataset}d and e)
is about 25 \kms, which is measured by the slope in the distance-time diagram
when the rain reaches the slit (1800 s $<$ t $<$ 2500 s) in Figure \ref{fig:vt_st}d.
This indicates that the rain drops at an absolute velocity of 90 \kms
and that the rain trajectory prior to impact is mostly directed along the line-of-sight.
In addition, we can see weaker upflows besides the strong downflows. We discuss the upflows in Section \ref{sec:discussion}.

\subsection{Temporal Variation of the IRIS Spectral Data}
There are two types of variability seen in the temporal variation of the IRIS spectra shown in Figures \ref{fig:vt_st}a--c.
One is the saw-tooth continuous variation, 
and the other are short-term bursts (Figures \ref{fig:vt_st}f--h).
The saw-tooth variation is seen over the whole observation period, 
whereas the bursty one is seen only when the rain reaches the IRIS slit (1800 s $\le$ t $\le$ 2500 s).
This indicates that the bursty variation has an association with the coronal rain.
In order to estimate the timescale of these saw-tooth and bursty variations, we did a wavelet analysis.

\begin{figure}[t]
\centerline{\includegraphics[width=13cm]{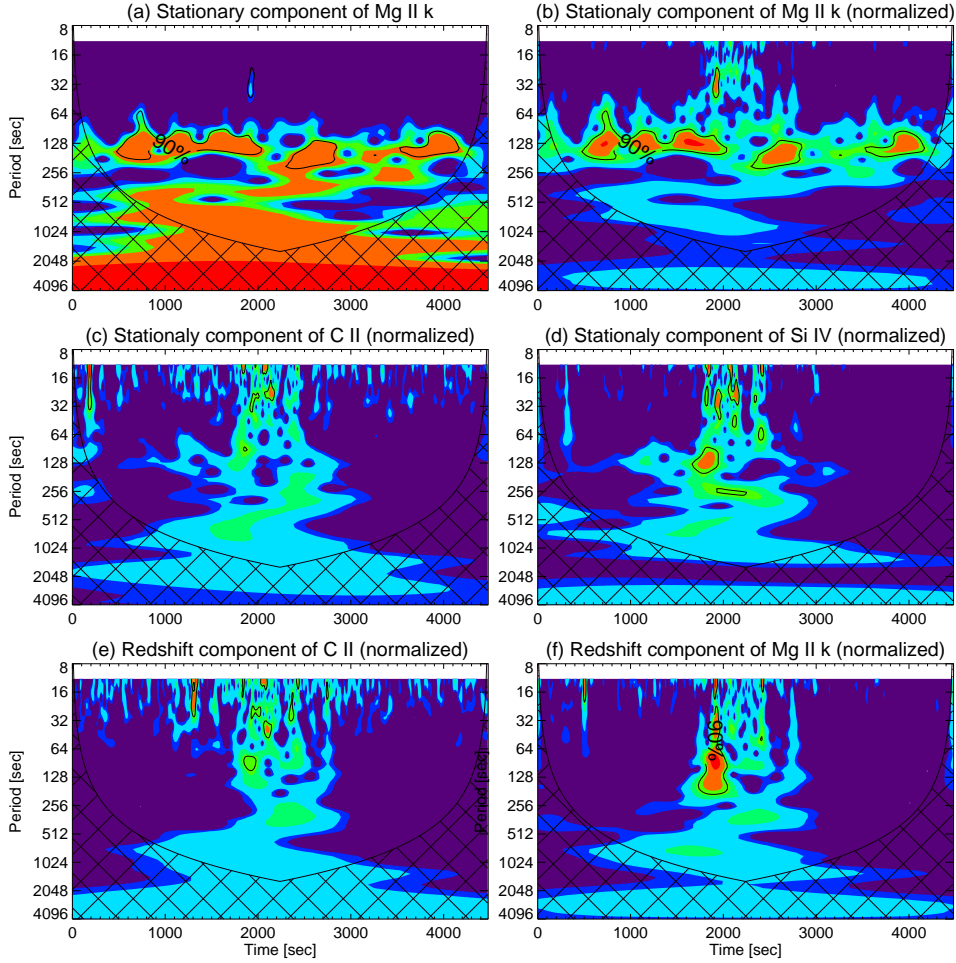}}
\caption{
(a) The wavelet power spectrum for the intensity variation of the stationary component of Mg {\sc ii} k.
(b)-(d) The wavelet power spectra for the intensity variation of the stationary component of Mg {\sc ii} k, C {\sc ii}, and Si {\sc iv}
normalized with $\sigma (\nu)$ defined in Equation \ref{eq:power_law_fit}.
(e)-(f) The wavelet power spectra for the intensity variation of the redshifted components of Mg {\sc ii} k and C {\sc ii}
normalized with $\sigma(\nu)$.
The black contours indicate the confidence level of 90\% computed with $\sigma(\nu)$.
The time-axis is the same as that of Figure \ref{fig:vt_st}.}
\label{fig:WPS_intens}
\end{figure}

\begin{figure}[t]
\centerline{\includegraphics[width=13cm]{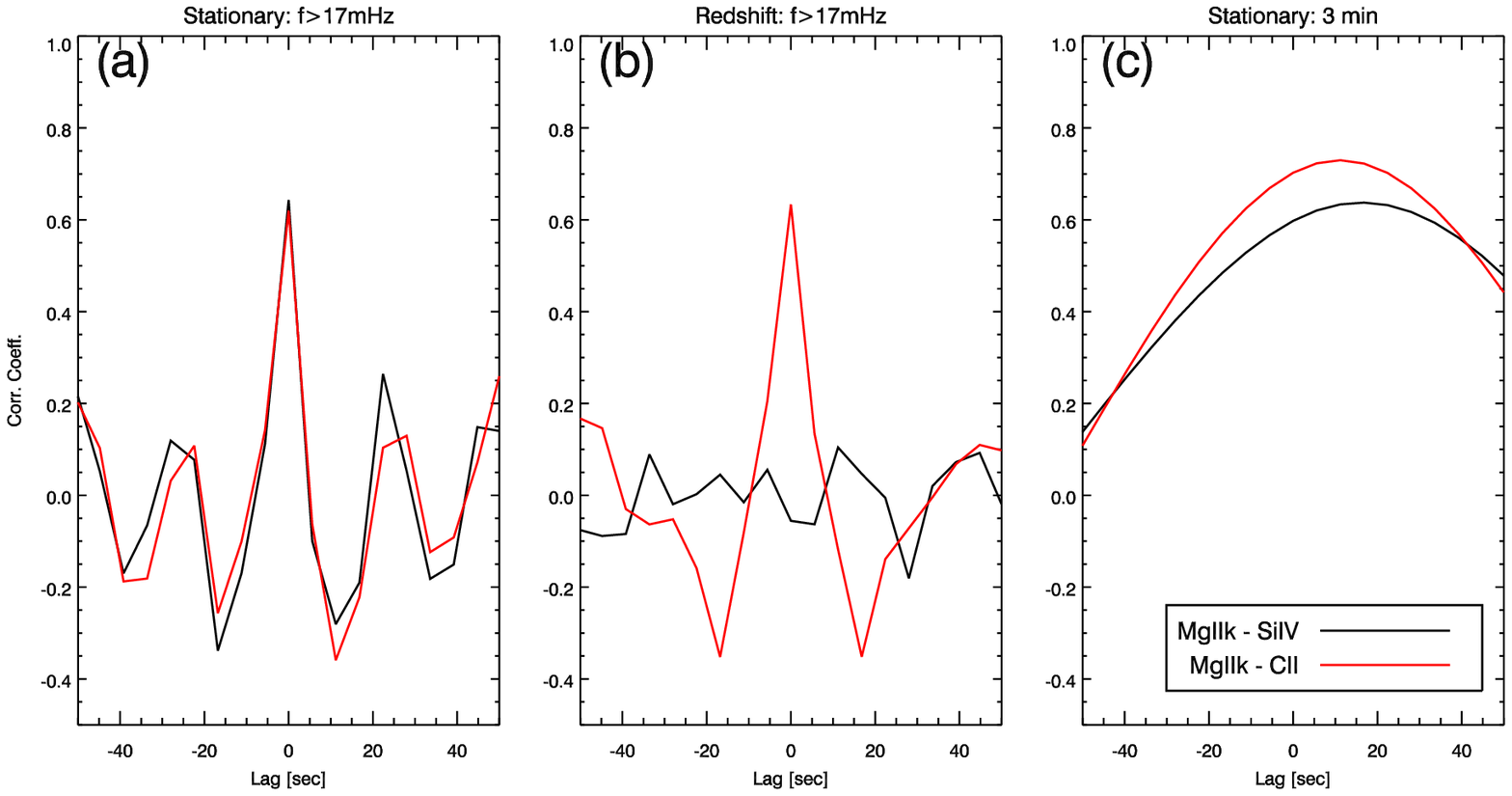}}
\caption{Wavelet power spectra of the stationary components of Mg {\sc ii} k (left) and Si {\sc iv} (right).
The blue solid lines show the time-averaged wavelet spectra and the blue dashed lines show the power law fits $\sigma(\nu)$ of them.
The black solid lines show the wavelet power spectra at t=1920 s and the red solid lines show the 90\% confidence levels
calculated with $\sigma(\nu)$.}
\label{fig:wps_slice}
\end{figure}

The time series of intensity variations of the stationary component
is derived by the total intensity
of a single gaussian fitting as shown by the blue line in Figure \ref{fig:vt_st}i.
The intensity variation of the redshifted components is defined by the total intensity of the residual 
described as the red region in Figure \ref{fig:vt_st}i.
The intensity variations of both the stationary and the redshifted components are normalized
with the noise levels measured at the end of the data (t $>$ 3900 s) as explained in \citet{Gabriel02}.
Figure \ref{fig:WPS_intens}a shows the wavelet power spectra for the intensity variation
of the stationary component of Mg {\sc ii} k line.
We computed the time-averaged wavelet spectra in the way developed in \citet{Auchere16} and fitted them with power law functions \citep{Kayshap20}:
\begin{equation}
\sigma (\nu) = A \nu^{s}, \label{eq:power_law_fit}
\end{equation}
where $A$, $\nu$, and $s$ are a coefficient, frequency, and a power law index, respectively.
The 90\% global confidence levels are calculated based on $\sigma (\nu)$ \citep{Auchere16}.
In Figure \ref{fig:wps_slice} we show the time-averaged wavelet spectra (blue solid lines), the power law fits (blue dashed lines),
the wavelet power spectra at t=1920 s (black solid lines), and the 90\% confidence levels (red solid lines) for Mg {\sc ii} k (left) and Si {\sc iv} (right).
The figure demonstrates how the power law function $\sigma(\nu)$ is obtained by fitting of the wavelet power spectra.
Figures \ref{fig:WPS_intens}b--d show the wavelet spectra for the stationary components
of Mg {\sc ii} k, C {\sc ii}, and Si {\sc iv} lines normalized with $\sigma(\nu)$,
and panels (e) and (f) are the same figures as panels (b) and (c) but for the intensity variation of the redshifted components.
There is a component with 3 minute periodicity all over the time series, which is clearly seen especially in the Mg {\sc ii} k line
(Figures \ref{fig:WPS_intens}a and b).
This corresponds to the 3 minute sunspot oscillation, which behaves as a shock wave 
as it propagates from the chromosphere to the TR, 
including impulsive blueshifts accompanied by intensity enhancements \citep{Tian14}.
On the other hand, the short-term component with a period shorter than 1 minute
corresponding to the bursty variation
has large power only in the period when the rain reaches the IRIS slit (from 1800 to 2500 s),
which may indicate that the short-term variation is excited by the coronal rain,
although the wavelet power spectra of the short-term component are broad.

\subsection{Time Lag of the Intensity Enhancements among the Lines}{\label{sec:timelag}}

\begin{figure}[t]
\centerline{\includegraphics[width=13cm]{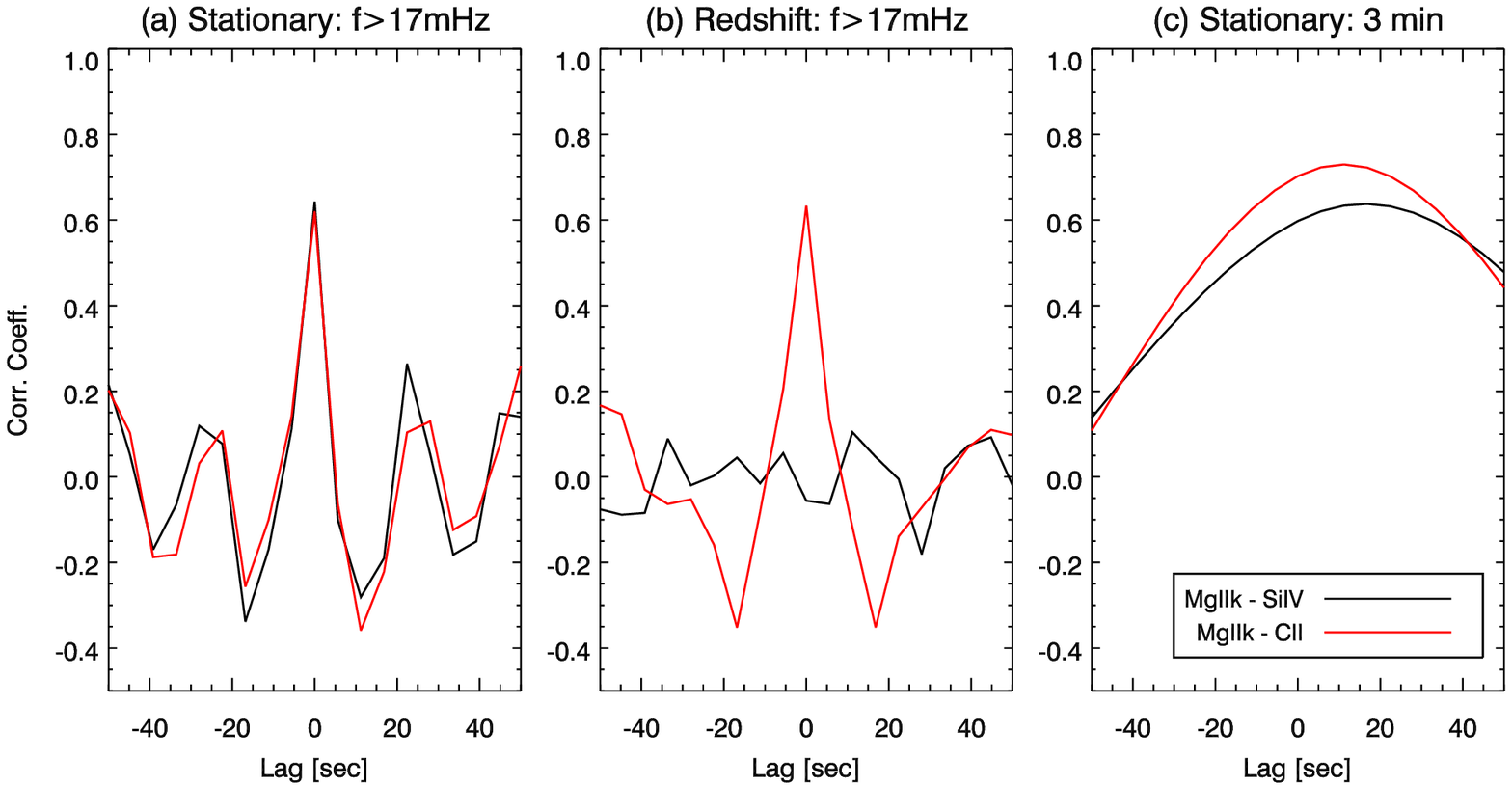}}
\caption{Cross-correlation coefficients between intensity variations of Mg {\sc ii} k
and those of Si {\sc iv} and  C {\sc ii} as a function of time lag in the rain time (from 1800 s to 2500 s), 
where positive time lag means the Mg {\sc ii} k line precedes; 
correlation functions (a) for short timescale intensity variations of the stationary components, 
(b) for short-term intensity variations with a timescale shorter than 1 min of the redshift components,
(c) for 3 minute intensity variations of the stationary components.
We use a high-pass filter with the cut-off frequency is around 17 mHz and with a cosine roll-off for (a) and (b)
and use a rectangle band-pass filter with a full width of 5.1 mHz for (c)
to extract specific timescale components from the original data.}
\label{fig:corr}
\end{figure}

For revealing the nature of intensity enhancements,
it is important to clarify the temporal order of
the intensity enhancements between the three IRIS lines.
We calculated the cross-correlation coefficients between the intensity variations of Mg {\sc ii} k and those of Si {\sc iv} and C {\sc ii}
as functions of time lag at each time-scale, as shown in Figure \ref{fig:corr}.
Since the IRIS spectral data were taken simultaneously by the three lines according to time stamps in the data,
we do not need interpolation of the time series for calculating cross-correlation coefficients between them.
As we see in Figure \ref{fig:WPS_intens},
there are two types of intensity variations:
the saw-tooth continuous variation with a periodicity of 3 minutes
and the short-term bursts with a timescale shorter than 1 minutes.
Therefore we calculated the cross-correlation coefficient
for each type of intensity variation extracted from the original intensity variation
with band-pass and high-pass filters.

Figure \ref{fig:corr}a shows the correlation functions of stationary components
derived with a high-pass filter that 
extracts the short-term variation from the original time-series.
We use a high-pass filter with the cut-off frequency around 17 mHz (corresponding to 1 minutes) and with a cosine roll-off.
The short-term variations of the stationary component have correlation coefficients
larger than 0.6 at a time lag of 0 seconds although determination of the peak time
lag is limited by the time cadence of the IRIS spectral data (5.6 sec).

Figure \ref{fig:corr}b is the same as the one in panel a but for the redshifted component.
The correlation coefficient between the intensity variations of Mg {\sc ii} k and C {\sc ii}
has large correlation coefficients like those seen in panel a,
whereas that between Mg {\sc ii} k and Si {\sc iv} has almost no correlation,
which is probably due to the low signal-to-noise ratio of the redshifted component of Si {\sc iv}.

Figure \ref{fig:corr}c is the same as that in panel a but processed with a band-pass filter that extracts 3 minutes variation.
The full width of this band-pass filter is 5.1 mHz in the frequency domain.
In contrast to the correlation coefficients of the short-term variation shown in Figures \ref{fig:corr}a and b,
the intensities of the IRIS lines are enhanced in the order of Mg {\sc ii} k, C {\sc ii}, and Si {\sc iv} in 3 minutes variation,
indicating that such enhancements occur in increasing temperature order, from the lower atmosphere.
The intensity enhancements in C {\sc ii} and Si {\sc iv} lag the Mg {\sc ii} k by 11.2 and 16.8 seconds, respectively.
Similar repeated patterns have been reported for dynamic fibrils in plages and
for sunspots \citep{Tian14} and they concluded that such patterns are driven by 
dynamic chromospheric shock waves.
The saw-tooth periodic structure seen in Figure \ref{fig:vt_st}a
consisting of impulsive blueshifts accompanied by gradual redshifts also supports this interpretation.

\section{Summary and Discussion}\label{sec:discussion}
In the previous section, we estimated the properties of the on-disk coronal rain moving towards the sunspot umbra.
The coronal rain clumps fall with Doppler velocity and plane-of-sky velocity of 85 \kms and 25 \kms, respectively, 
which yields an actual velocity of 90 \kms, i.e., much faster than the local sound speeds in the TR and the chromosphere
since the local sound speed is $10-50$ \kms when the temperature is $10^4-10^5$ K
if we use an adiabatic sound speed of $\sqrt{\gamma k_B T/m}$
where $\gamma$, $k_B$, $T$, and $m$ are the heat capacity ratio, the Boltzmann constant, temperature, and mass of a particle, respectively.
As shown in Figures \ref{fig:corr}a and b, there are no significant time-lags
between the short-term intensity variations of the three IRIS lines.
These results indicate that all the lines are heated up concurrently,
most probably due to shock waves induced by the supersonic coronal rain.
This local heating interpretation is also consistent with the weaker upflows
besides the strong downflows seen in Figure \ref{fig:vt_st}.
These weaker upflows suggest that mild evaporation is produced by the heating by the coronal rain
besides the strong downflows seen at the Mg {\sc ii} k line in Figure \ref{fig:vt_st}f \citep{Kleint14}.
The coronal rain clumps have a width of 580 km and show rapid intensity fluctuations with a timescale shorter than 1 minute.
Hereafter we use typical timescale of the coronal rain of 30 seconds.
If we multiply the timescale of 30 sec by the velocity of 90 \kms, the typical length of the clumps is estimated to be about $2.7$ Mm.
Such widths and lengths are consistent with those observed in the same lines and instrument by \citet{Antolin15}.
The temperature should be about $10^5$ K equivalent to the formation temperature of the Si {\sc iv} line.

The clear relationship between the apparent flows falling onto the sunspot umbra and the intensity enhancements
of stationary components with strong Doppler shift
suggest that the coronal rain induces the local heating events.
These heating events are possibly due to the shock made by the supersonic downflows of the coronal rain.
The general scenario of the local heating by the coronal rain is the same as in previous reports \citep{Kleint14, Reale13}. 
However, the location of the shock front, or equivalently the height of the plasma heating, remains unclear.
The falling material can penetrate the atmosphere to a dense layer whose gas pressure is equal to the ram pressure of the rain. 
The ram pressure $P_{ram} = \frac{1}{2} \rho v^2$ is estimated about $9.5 ~\mathrm{dyn~cm^{-2}}$,
where we use observed velocity 90 \kms and previously reported average electron density 
$1.4 \times 10^{11} ~\mathrm{cm^{-3}}$ \citep{Antolin15}.

When the coronal rain collides with the chromospheric plasma, a part of the kinetic energy of coronal rain is 
converted into thermal energy, heating chromospheric plasma and the falling coronal rain.
We can estimate the amount of the kinetic energy of a clump of the coronal rain $E_{clump}$ to be
\begin{equation}
E_{clump} = \frac{1}{2} \rho v^2 V_{clump} = \frac{1}{2} m_p n_e v^2 \left( \pi r^2 l \right), \label{eq:Eclump}
\end{equation}
where $V_{clump}$ is the volume of a clump, and we approximate its shape by a cylinder 
with the radius $r$ and the length $l$.
In this case, the radius is equivalent to the half of the clump's width of $580 ~\mathrm{km}$ corresponding to 5 IRIS pixels.
We use the electron density $1.4 \times 10^{11} ~\mathrm{cm^{-3}}$.
If we apply the observed values, $r=290 ~\mathrm{km}$, $l=2.7 ~\mathrm{Mm}$, and $v=90 ~\mathrm{km\ s^{-1}}$,
then we get the kinetic energy of about $6.8 \times 10^{24}$ erg.
Let us assume the condition that the number density in the chromosphere is $10^{12} ~\mathrm{cm^{-3}}$ and
the volume is $\pi r^2 H$, where $H=300 ~\mathrm{km}$ is the pressure scale height of the chromosphere.
If the temperature increase from $\log T=4$ to $\log T=5$, 
the required heating energy $nk_B \Delta T (\pi r^2 H)$ is estimated to be about $1 \times 10^{24}$ erg.
Therefore, by converting a fraction of the kinetic energy, the coronal rain can heat the chromosphere.

The most important question is what produces such small-scale structures.
When a coronal rain clump is first produced in the corona,
the typical size of the clump may reflect the spatial scales of dynamical processes within the loops.
In principle, a condensation will form and accumulate plasma as long as a pressure gradient exists. 
This pressure gradient is created initially by the loss of pressure from thermal instability. 
During its precipitation from the corona to the lower layers, the coronal rain clump could be broken 
into a smaller pieces by the Kelvin-Helmholtz instability (KHI) 
excited by the velocity shear between the clump and the surrounding stationary atmosphere.
However, if the velocity vector of the coronal rain clump is parallel to the magnetic field,
the KHI may not grow easily since the magnetic tension is likely to suppress KHI vortices.
In 2.5-D MHD simulations by \citet{Fang15} both very long and short clumps can be formed.
Although their simulation does not provide the sufficient resolution to investigate the formation of the KHI,
they discuss how the shear flows could lead to the break-up of the clumps, 
and they suggest a possible influence of the KHI. 
On the other hand, \citet{MartinezGomez19} performed two-dimentional ideal MHD simulation of coronal rain
and concluded that the KHI are not expected to develop in typical coronal rain conditions.
They argued that the length of coronal rain may be explained by the timescale of the catastrophic cooling.
Future high-resolution and multi-wavelength observations such as
Solar-C\_EUV High-Throughput Spectroscopic Telescope (EUVST: \citealt{Shimizu+19})
will reveal the formation mechanism of clumps in coronal rain.

%

%
\begin{acks}
Data are a courtesy of the science teams, IRIS and SDO.
IRIS is a NASA small explorer mission developed and operated by LMSAL
with mission operations executed at NASA Ames Research
center and major contributions to downlink communications
funded by ESA and the Norwegian Space Centre.
AIA is an instrument on board SDO, a mission for NASA's Living with a Star program.
R.T.I. is supported by JSPS Research Fellowships for Young Scientists.
P.A. acknowledges funding from his STFC Ernest Rutherford Fellowship (No. ST/R004285/1).
This work was supported by JSPS KAKENHI Grant Numbers JP16K17671 (PI: S. Toriumi), 
JP15H05814 (PI: K. Ichimoto), JP25220703 (PI: S. Tsuneta), JP18H05234 (PI: Y. Katsukawa), and JP19J20294 (PI: R.T. Ishikawa).
\end{acks}

\footnotesize\paragraph*{Disclosure of Potential Conflicts of Interest}
The authors declare that there are no conflicts of interest.

%
%
\bibliographystyle{spr-mp-sola}
\bibliography{reference}  

\begin{thebibliography}{32}
\ifx\bisbn     \undefined \def\bisbn  #1{ISBN #1}\fi
\ifx\binits    \undefined \def\binits#1{#1}\fi
\ifx\bauthor   \undefined \def\bauthor#1{#1}\fi
\ifx\batitle   \undefined \def\batitle#1{#1}\fi
\ifx\bjtitle   \undefined \def\bjtitle#1{\textit{#1}}\fi
\ifx\bvolume   \undefined \def\bvolume#1{\textbf{#1}}\fi
\ifx\byear     \undefined \def\byear#1{#1}\fi
\ifx\bissue    \undefined \def\bissue#1{#1}\fi
\ifx\bfpage    \undefined \def\bfpage#1{#1}\fi
\ifx\blpage    \undefined \def\blpage #1{#1}\fi
\ifx\burl      \undefined \def\burl#1{\textsf{#1}}\fi
\ifx\href      \undefined \def\href#1#2{\textsf{#2}}\fi
\ifx\betal     \undefined \def\betal{\textit{et al.}}\fi
\ifx\bctitle   \undefined \def\bctitle#1{#1}\fi
\ifx\beditor   \undefined \def\beditor#1{#1}\fi
\ifx\bbtitle   \undefined \def\bbtitle#1{\textit{#1}}\fi
\ifx\bedition  \undefined \def\bedition#1{#1}\fi
\ifx\bseriesno \undefined \def\bseriesno#1{\textbf{#1}}\fi
\ifx\blocation \undefined \def\blocation#1{#1}\fi
\ifx\bsertitle \undefined \def\bsertitle#1{\textit{#1}}\fi
\ifx\bsnm      \undefined \def\bsnm#1{#1}\fi
\ifx\bsuffix   \undefined \def\bsuffix#1{#1}\fi
\ifx\bparticle \undefined \def\bparticle#1{#1}\fi
\ifx\barticle  \undefined \def\barticle#1{}\fi
\ifx\binstitute  \undefined \def\binstitute#1{#1}\fi
\ifx\bpublisher  \undefined \def\bpublisher#1{#1}\fi
\ifx\doiurl    \undefined \def\doiurl#1{\href{#1}{\textsf{DOI}}}\fi
\makeatletter
\def\safeHref#1#2#3{\in@{http}{#2}\ifin@\href{#2}{#3}\else\href{#1#2}{#3}\fi}
\makeatother
\ifx\adsurl    \undefined
  \def\adsurl#1{\safeHref{https://ui.adsabs.harvard.edu/abs/}{#1}{\textsf{ADS}}}\fi
\ifx\arxivurl  \undefined
  \def\arxivurl#1{\safeHref{http://arxiv.org/abs/}{#1}{\textsf{arXiv}}}\fi
\ifx\botherref \undefined \def\botherref#1{}\fi
\ifx\url       \undefined \def\url#1{\textsf{#1}}\fi
\ifx\bchapter  \undefined \def\bchapter#1{}\fi
\ifx\bbook     \undefined \def\bbook#1{}\fi
\ifx\bcomment  \undefined \def\bcomment#1{#1}\fi
\ifx\oauthor   \undefined \def\oauthor#1{#1}\fi
\ifx\citeauthoryear \undefined\def \citeauthoryear#1{#1}\fi
\def\endbibitem {}
\ifx\bconflocation  \undefined \def\bconflocation#1{#1} \fi

\bibitem[\protect\citeauthoryear{{Antiochos} and
  {Klimchuk}}{1991}]{Antiochos&Klimchuk91}
\begin{barticle}
\bauthor{\bsnm{{Antiochos}}, \binits{S.K.}},
\bauthor{\bsnm{{Klimchuk}}, \binits{J.A.}}:
\byear{1991},
\batitle{{A model for the formation of solar prominences}}.
\bjtitle{\apj}
\bvolume{378},
\bfpage{372}.
\doiurl{https://doi.org/10.1086/170437}.
\adsurl{1991ApJ...378..372A}.
\end{barticle}
\endbibitem

\bibitem[\protect\citeauthoryear{{Antolin} and {Rouppe van der
  Voort}}{2012}]{Antolin12a}
\begin{barticle}
\bauthor{\bsnm{{Antolin}}, \binits{P.}},
\bauthor{\bsnm{{Rouppe van der Voort}}, \binits{L.}}:
\byear{2012},
\batitle{{Observing the Fine Structure of Loops through High-resolution
  Spectroscopic Observations of Coronal Rain with the CRISP Instrument at the
  Swedish Solar Telescope}}.
\bjtitle{\apj}
\bvolume{745},
\bfpage{152}.
\doiurl{https://doi.org/10.1088/0004-637X/745/2/152}.
\adsurl{2012ApJ...745..152A}.
\end{barticle}
\endbibitem

\bibitem[\protect\citeauthoryear{{Antolin}, {Shibata}, and
  {Vissers}}{2010}]{Antolin10}
\begin{barticle}
\bauthor{\bsnm{{Antolin}}, \binits{P.}},
\bauthor{\bsnm{{Shibata}}, \binits{K.}},
\bauthor{\bsnm{{Vissers}}, \binits{G.}}:
\byear{2010},
\batitle{{Coronal Rain as a Marker for Coronal Heating Mechanisms}}.
\bjtitle{\apj}
\bvolume{716},
\bfpage{154}.
\doiurl{https://doi.org/10.1088/0004-637X/716/1/154}.
\adsurl{2010ApJ...716..154A}.
\end{barticle}
\endbibitem

\bibitem[\protect\citeauthoryear{{Antolin}, {Vissers}, and {Rouppe van der
  Voort}}{2012}]{Antolin12b}
\begin{barticle}
\bauthor{\bsnm{{Antolin}}, \binits{P.}},
\bauthor{\bsnm{{Vissers}}, \binits{G.}},
\bauthor{\bsnm{{Rouppe van der Voort}}, \binits{L.}}:
\byear{2012},
\batitle{{On-Disk Coronal Rain}}.
\bjtitle{\solphys}
\bvolume{280},
\bfpage{457}.
\doiurl{https://doi.org/10.1007/s11207-012-9979-7}.
\adsurl{2012SoPh..280..457A}.
\end{barticle}
\endbibitem

\bibitem[\protect\citeauthoryear{{Antolin} \textit{et~al.}}{2015}]{Antolin15}
\begin{barticle}
\bauthor{\bsnm{{Antolin}}, \binits{P.}},
\bauthor{\bsnm{{Vissers}}, \binits{G.}},
\bauthor{\bsnm{{Pereira}}, \binits{T.M.D.}},
\bauthor{\bsnm{{Rouppe van der Voort}}, \binits{L.}},
\bauthor{\bsnm{{Scullion}}, \binits{E.}}:
\byear{2015},
\batitle{{The Multithermal and Multi-stranded Nature of Coronal Rain}}.
\bjtitle{\apj}
\bvolume{806},
\bfpage{81}.
\doiurl{https://doi.org/10.1088/0004-637X/806/1/81}.
\adsurl{2015ApJ...806...81A}.
\end{barticle}
\endbibitem

\bibitem[\protect\citeauthoryear{{Auch{\`e}re}
  \textit{et~al.}}{2014}]{Auchere14}
\begin{barticle}
\bauthor{\bsnm{{Auch{\`e}re}}, \binits{F.}},
\bauthor{\bsnm{{Bocchialini}}, \binits{K.}},
\bauthor{\bsnm{{Solomon}}, \binits{J.}},
\bauthor{\bsnm{{Tison}}, \binits{E.}}:
\byear{2014},
\batitle{{Long-period intensity pulsations in the solar corona during activity
  cycle 23}}.
\bjtitle{\aap}
\bvolume{563},
\bfpage{A8}.
\doiurl{https://doi.org/10.1051/0004-6361/201322572}.
\adsurl{2014A\%26A...563A...8A}.
\end{barticle}
\endbibitem

\bibitem[\protect\citeauthoryear{{Auch{\`e}re}
  \textit{et~al.}}{2016}]{Auchere16}
\begin{barticle}
\bauthor{\bsnm{{Auch{\`e}re}}, \binits{F.}},
\bauthor{\bsnm{{Froment}}, \binits{C.}},
\bauthor{\bsnm{{Bocchialini}}, \binits{K.}},
\bauthor{\bsnm{{Buchlin}}, \binits{E.}},
\bauthor{\bsnm{{Solomon}}, \binits{J.}}:
\byear{2016},
\batitle{{On the Fourier and Wavelet Analysis of Coronal Time Series}}.
\bjtitle{\apj}
\bvolume{825}(\bissue{2}),
\bfpage{110}.
\doiurl{https://doi.org/10.3847/0004-637X/825/2/110}.
\adsurl{2016ApJ...825..110A}.
\end{barticle}
\endbibitem

\bibitem[\protect\citeauthoryear{{de Groof} \textit{et~al.}}{2005}]{DeGroof05}
\begin{barticle}
\bauthor{\bsnm{{de Groof}}, \binits{A.}},
\bauthor{\bsnm{{Bastiaensen}}, \binits{C.}},
\bauthor{\bsnm{{M{\"u}ller}}, \binits{D.A.N.}},
\bauthor{\bsnm{{Berghmans}}, \binits{D.}},
\bauthor{\bsnm{{Poedts}}, \binits{S.}}:
\byear{2005},
\batitle{{Detailed comparison of downflows seen both in EIT 30.4 nm and Big
  Bear H{$\alpha$} movies}}.
\bjtitle{\aap}
\bvolume{443},
\bfpage{319}.
\doiurl{https://doi.org/10.1051/0004-6361:20053129}.
\adsurl{2005A\%26A...443..319D}.
\end{barticle}
\endbibitem

\bibitem[\protect\citeauthoryear{{De Pontieu}
  \textit{et~al.}}{2014}]{DePontieu14}
\begin{barticle}
\bauthor{\bsnm{{De Pontieu}}, \binits{B.}},
\bauthor{\bsnm{{Title}}, \binits{A.M.}},
\bauthor{\bsnm{{Lemen}}, \binits{J.R.}},
\bauthor{\bsnm{{Kushner}}, \binits{G.D.}},
\bauthor{\bsnm{{Akin}}, \binits{D.J.}},
\bauthor{\bsnm{{Allard}}, \binits{B.}},
\bauthor{\bsnm{{Berger}}, \binits{T.}},
\bauthor{\bsnm{{Boerner}}, \binits{P.}},
\bauthor{\bsnm{{Cheung}}, \binits{M.}},
\bauthor{\bsnm{{Chou}}, \binits{C.}},
\bauthor{\bsnm{{Drake}}, \binits{J.F.}},
\bauthor{\bsnm{{Duncan}}, \binits{D.W.}},
\bauthor{\bsnm{{Freeland}}, \binits{S.}},
\bauthor{\bsnm{{Heyman}}, \binits{G.F.}},
\bauthor{\bsnm{{Hoffman}}, \binits{C.}},
\bauthor{\bsnm{{Hurlburt}}, \binits{N.E.}},
\bauthor{\bsnm{{Lindgren}}, \binits{R.W.}},
\bauthor{\bsnm{{Mathur}}, \binits{D.}},
\bauthor{\bsnm{{Rehse}}, \binits{R.}},
\bauthor{\bsnm{{Sabolish}}, \binits{D.}},
\bauthor{\bsnm{{Seguin}}, \binits{R.}},
\bauthor{\bsnm{{Schrijver}}, \binits{C.J.}},
\bauthor{\bsnm{{Tarbell}}, \binits{T.D.}},
\bauthor{\bsnm{{W{\"u}lser}}, \binits{J.-P.}},
\bauthor{\bsnm{{Wolfson}}, \binits{C.J.}},
\bauthor{\bsnm{{Yanari}}, \binits{C.}},
\bauthor{\bsnm{{Mudge}}, \binits{J.}},
\bauthor{\bsnm{{Nguyen-Phuc}}, \binits{N.}},
\bauthor{\bsnm{{Timmons}}, \binits{R.}},
\bauthor{\bsnm{{van Bezooijen}}, \binits{R.}},
\bauthor{\bsnm{{Weingrod}}, \binits{I.}},
\bauthor{\bsnm{{Brookner}}, \binits{R.}},
\bauthor{\bsnm{{Butcher}}, \binits{G.}},
\bauthor{\bsnm{{Dougherty}}, \binits{B.}},
\bauthor{\bsnm{{Eder}}, \binits{J.}},
\bauthor{\bsnm{{Knagenhjelm}}, \binits{V.}},
\bauthor{\bsnm{{Larsen}}, \binits{S.}},
\bauthor{\bsnm{{Mansir}}, \binits{D.}},
\bauthor{\bsnm{{Phan}}, \binits{L.}},
\bauthor{\bsnm{{Boyle}}, \binits{P.}},
\bauthor{\bsnm{{Cheimets}}, \binits{P.N.}},
\bauthor{\bsnm{{DeLuca}}, \binits{E.E.}},
\bauthor{\bsnm{{Golub}}, \binits{L.}},
\bauthor{\bsnm{{Gates}}, \binits{R.}},
\bauthor{\bsnm{{Hertz}}, \binits{E.}},
\bauthor{\bsnm{{McKillop}}, \binits{S.}},
\bauthor{\bsnm{{Park}}, \binits{S.}},
\bauthor{\bsnm{{Perry}}, \binits{T.}},
\bauthor{\bsnm{{Podgorski}}, \binits{W.A.}},
\bauthor{\bsnm{{Reeves}}, \binits{K.}},
\bauthor{\bsnm{{Saar}}, \binits{S.}},
\bauthor{\bsnm{{Testa}}, \binits{P.}},
\bauthor{\bsnm{{Tian}}, \binits{H.}},
\bauthor{\bsnm{{Weber}}, \binits{M.}},
\bauthor{\bsnm{{Dunn}}, \binits{C.}},
\bauthor{\bsnm{{Eccles}}, \binits{S.}},
\bauthor{\bsnm{{Jaeggli}}, \binits{S.A.}},
\bauthor{\bsnm{{Kankelborg}}, \binits{C.C.}},
\bauthor{\bsnm{{Mashburn}}, \binits{K.}},
\bauthor{\bsnm{{Pust}}, \binits{N.}},
\bauthor{\bsnm{{Springer}}, \binits{L.}},
\bauthor{\bsnm{{Carvalho}}, \binits{R.}},
\bauthor{\bsnm{{Kleint}}, \binits{L.}},
\bauthor{\bsnm{{Marmie}}, \binits{J.}},
\bauthor{\bsnm{{Mazmanian}}, \binits{E.}},
\bauthor{\bsnm{{Pereira}}, \binits{T.M.D.}},
\bauthor{\bsnm{{Sawyer}}, \binits{S.}},
\bauthor{\bsnm{{Strong}}, \binits{J.}},
\bauthor{\bsnm{{Worden}}, \binits{S.P.}},
\bauthor{\bsnm{{Carlsson}}, \binits{M.}},
\bauthor{\bsnm{{Hansteen}}, \binits{V.H.}},
\bauthor{\bsnm{{Leenaarts}}, \binits{J.}},
\bauthor{\bsnm{{Wiesmann}}, \binits{M.}},
\bauthor{\bsnm{{Aloise}}, \binits{J.}},
\bauthor{\bsnm{{Chu}}, \binits{K.-C.}},
\bauthor{\bsnm{{Bush}}, \binits{R.I.}},
\bauthor{\bsnm{{Scherrer}}, \binits{P.H.}},
\bauthor{\bsnm{{Brekke}}, \binits{P.}},
\bauthor{\bsnm{{Martinez-Sykora}}, \binits{J.}},
\bauthor{\bsnm{{Lites}}, \binits{B.W.}},
\bauthor{\bsnm{{McIntosh}}, \binits{S.W.}},
\bauthor{\bsnm{{Uitenbroek}}, \binits{H.}},
\bauthor{\bsnm{{Okamoto}}, \binits{T.J.}},
\bauthor{\bsnm{{Gummin}}, \binits{M.A.}},
\bauthor{\bsnm{{Auker}}, \binits{G.}},
\bauthor{\bsnm{{Jerram}}, \binits{P.}},
\bauthor{\bsnm{{Pool}}, \binits{P.}},
\bauthor{\bsnm{{Waltham}}, \binits{N.}}:
\byear{2014},
\batitle{{The Interface Region Imaging Spectrograph (IRIS)}}.
\bjtitle{\solphys}
\bvolume{289},
\bfpage{2733}.
\doiurl{https://doi.org/10.1007/s11207-014-0485-y}.
\adsurl{2014SoPh..289.2733D}.
\end{barticle}
\endbibitem

\bibitem[\protect\citeauthoryear{{Fang}, {Xia}, and {Keppens}}{2013}]{Fang13}
\begin{barticle}
\bauthor{\bsnm{{Fang}}, \binits{X.}},
\bauthor{\bsnm{{Xia}}, \binits{C.}},
\bauthor{\bsnm{{Keppens}}, \binits{R.}}:
\byear{2013},
\batitle{{Multidimensional Modeling of Coronal Rain Dynamics}}.
\bjtitle{\apjl}
\bvolume{771},
\bfpage{L29}.
\doiurl{https://doi.org/10.1088/2041-8205/771/2/L29}.
\adsurl{2013ApJ...771L..29F}.
\end{barticle}
\endbibitem

\bibitem[\protect\citeauthoryear{{Fang} \textit{et~al.}}{2015}]{Fang15}
\begin{barticle}
\bauthor{\bsnm{{Fang}}, \binits{X.}},
\bauthor{\bsnm{{Xia}}, \binits{C.}},
\bauthor{\bsnm{{Keppens}}, \binits{R.}},
\bauthor{\bsnm{{Van Doorsselaere}}, \binits{T.}}:
\byear{2015},
\batitle{{Coronal Rain in Magnetic Arcades: Rebound Shocks, Limit Cycles, and
  Shear Flows}}.
\bjtitle{\apj}
\bvolume{807},
\bfpage{142}.
\doiurl{https://doi.org/10.1088/0004-637X/807/2/142}.
\adsurl{2015ApJ...807..142F}.
\end{barticle}
\endbibitem

\bibitem[\protect\citeauthoryear{{Froment} \textit{et~al.}}{2015}]{Froment15}
\begin{barticle}
\bauthor{\bsnm{{Froment}}, \binits{C.}},
\bauthor{\bsnm{{Auch{\`e}re}}, \binits{F.}},
\bauthor{\bsnm{{Bocchialini}}, \binits{K.}},
\bauthor{\bsnm{{Buchlin}}, \binits{E.}},
\bauthor{\bsnm{{Guennou}}, \binits{C.}},
\bauthor{\bsnm{{Solomon}}, \binits{J.}}:
\byear{2015},
\batitle{{Evidence for Evaporation-incomplete Condensation Cycles in Warm Solar
  Coronal Loops}}.
\bjtitle{\apj}
\bvolume{807},
\bfpage{158}.
\doiurl{https://doi.org/10.1088/0004-637X/807/2/158}.
\adsurl{2015ApJ...807..158F}.
\end{barticle}
\endbibitem

\bibitem[\protect\citeauthoryear{{Froment} \textit{et~al.}}{2017}]{Froment17}
\begin{barticle}
\bauthor{\bsnm{{Froment}}, \binits{C.}},
\bauthor{\bsnm{{Auch{\`e}re}}, \binits{F.}},
\bauthor{\bsnm{{Aulanier}}, \binits{G.}},
\bauthor{\bsnm{{Miki{\'c}}}, \binits{Z.}},
\bauthor{\bsnm{{Bocchialini}}, \binits{K.}},
\bauthor{\bsnm{{Buchlin}}, \binits{E.}},
\bauthor{\bsnm{{Solomon}}, \binits{J.}}:
\byear{2017},
\batitle{{Long-period Intensity Pulsations in Coronal Loops Explained by
  Thermal Non-equilibrium Cycles}}.
\bjtitle{\apj}
\bvolume{835},
\bfpage{272}.
\doiurl{https://doi.org/10.3847/1538-4357/835/2/272}.
\adsurl{2017ApJ...835..272F}.
\end{barticle}
\endbibitem

\bibitem[\protect\citeauthoryear{{Gabriel} \textit{et~al.}}{2002}]{Gabriel02}
\begin{barticle}
\bauthor{\bsnm{{Gabriel}}, \binits{A.H.}},
\bauthor{\bsnm{{Baudin}}, \binits{F.}},
\bauthor{\bsnm{{Boumier}}, \binits{P.}},
\bauthor{\bsnm{{Garc{\'\i}a}}, \binits{R.A.}},
\bauthor{\bsnm{{Turck-Chi{\`e}ze}}, \binits{S.}},
\bauthor{\bsnm{{Appourchaux}}, \binits{T.}},
\bauthor{\bsnm{{Bertello}}, \binits{L.}},
\bauthor{\bsnm{{Berthomieu}}, \binits{G.}},
\bauthor{\bsnm{{Charra}}, \binits{J.}},
\bauthor{\bsnm{{Gough}}, \binits{D.O.}},
\bauthor{\bsnm{{Pall{\'e}}}, \binits{P.L.}},
\bauthor{\bsnm{{Provost}}, \binits{J.}},
\bauthor{\bsnm{{Renaud}}, \binits{C.}},
\bauthor{\bsnm{{Robillot}}, \binits{J.-M.}},
\bauthor{\bsnm{{Roca Cort{\'e}s}}, \binits{T.}},
\bauthor{\bsnm{{Thiery}}, \binits{S.}},
\bauthor{\bsnm{{Ulrich}}, \binits{R.K.}}:
\byear{2002},
\batitle{{A search for solar g modes in the GOLF data}}.
\bjtitle{\aap}
\bvolume{390},
\bfpage{1119}.
\doiurl{https://doi.org/10.1051/0004-6361:20020695}.
\adsurl{2002A&A...390.1119G}.
\end{barticle}
\endbibitem

\bibitem[\protect\citeauthoryear{{Kawaguchi}}{1970}]{Kawaguchi70}
\begin{barticle}
\bauthor{\bsnm{{Kawaguchi}}, \binits{I.}}:
\byear{1970},
\batitle{{Observed Interaction between Prominences}}.
\bjtitle{\pasj}
\bvolume{22},
\bfpage{405}.
\adsurl{1970PASJ...22..405K}.
\end{barticle}
\endbibitem

\bibitem[\protect\citeauthoryear{{Kayshap} \textit{et~al.}}{2020}]{Kayshap20}
\begin{barticle}
\bauthor{\bsnm{{Kayshap}}, \binits{P.}},
\bauthor{\bsnm{{Srivastava}}, \binits{A.K.}},
\bauthor{\bsnm{{Tiwari}}, \binits{S.K.}},
\bauthor{\bsnm{{Jel{\'\i}nek}}, \binits{P.}},
\bauthor{\bsnm{{Mathioudakis}}, \binits{M.}}:
\byear{2020},
\batitle{{Propagation of waves above a plage as observed by IRIS and SDO}}.
\bjtitle{\aap}
\bvolume{634},
\bfpage{A63}.
\doiurl{https://doi.org/10.1051/0004-6361/201936070}.
\adsurl{2020A&A...634A..63K}.
\end{barticle}
\endbibitem

\bibitem[\protect\citeauthoryear{{Kleint} \textit{et~al.}}{2014}]{Kleint14}
\begin{barticle}
\bauthor{\bsnm{{Kleint}}, \binits{L.}},
\bauthor{\bsnm{{Antolin}}, \binits{P.}},
\bauthor{\bsnm{{Tian}}, \binits{H.}},
\bauthor{\bsnm{{Judge}}, \binits{P.}},
\bauthor{\bsnm{{Testa}}, \binits{P.}},
\bauthor{\bsnm{{De Pontieu}}, \binits{B.}},
\bauthor{\bsnm{{Mart{\'{\i}}nez-Sykora}}, \binits{J.}},
\bauthor{\bsnm{{Reeves}}, \binits{K.K.}},
\bauthor{\bsnm{{Wuelser}}, \binits{J.P.}},
\bauthor{\bsnm{{McKillop}}, \binits{S.}},
\bauthor{\bsnm{{Saar}}, \binits{S.}},
\bauthor{\bsnm{{Carlsson}}, \binits{M.}},
\bauthor{\bsnm{{Boerner}}, \binits{P.}},
\bauthor{\bsnm{{Hurlburt}}, \binits{N.}},
\bauthor{\bsnm{{Lemen}}, \binits{J.}},
\bauthor{\bsnm{{Tarbell}}, \binits{T.D.}},
\bauthor{\bsnm{{Title}}, \binits{A.}},
\bauthor{\bsnm{{Golub}}, \binits{L.}},
\bauthor{\bsnm{{Hansteen}}, \binits{V.}},
\bauthor{\bsnm{{Jaeggli}}, \binits{S.}},
\bauthor{\bsnm{{Kankelborg}}, \binits{C.}}:
\byear{2014},
\batitle{{Detection of Supersonic Downflows and Associated Heating Events in
  the Transition Region above Sunspots}}.
\bjtitle{\apjl}
\bvolume{789},
\bfpage{L42}.
\doiurl{https://doi.org/10.1088/2041-8205/789/2/L42}.
\adsurl{2014ApJ...789L..42K}.
\end{barticle}
\endbibitem

\bibitem[\protect\citeauthoryear{{Lemen} \textit{et~al.}}{2012}]{Lemen12}
\begin{barticle}
\bauthor{\bsnm{{Lemen}}, \binits{J.R.}},
\bauthor{\bsnm{{Title}}, \binits{A.M.}},
\bauthor{\bsnm{{Akin}}, \binits{D.J.}},
\bauthor{\bsnm{{Boerner}}, \binits{P.F.}},
\bauthor{\bsnm{{Chou}}, \binits{C.}},
\bauthor{\bsnm{{Drake}}, \binits{J.F.}},
\bauthor{\bsnm{{Duncan}}, \binits{D.W.}},
\bauthor{\bsnm{{Edwards}}, \binits{C.G.}},
\bauthor{\bsnm{{Friedlaender}}, \binits{F.M.}},
\bauthor{\bsnm{{Heyman}}, \binits{G.F.}},
\bauthor{\bsnm{{Hurlburt}}, \binits{N.E.}},
\bauthor{\bsnm{{Katz}}, \binits{N.L.}},
\bauthor{\bsnm{{Kushner}}, \binits{G.D.}},
\bauthor{\bsnm{{Levay}}, \binits{M.}},
\bauthor{\bsnm{{Lindgren}}, \binits{R.W.}},
\bauthor{\bsnm{{Mathur}}, \binits{D.P.}},
\bauthor{\bsnm{{McFeaters}}, \binits{E.L.}},
\bauthor{\bsnm{{Mitchell}}, \binits{S.}},
\bauthor{\bsnm{{Rehse}}, \binits{R.A.}},
\bauthor{\bsnm{{Schrijver}}, \binits{C.J.}},
\bauthor{\bsnm{{Springer}}, \binits{L.A.}},
\bauthor{\bsnm{{Stern}}, \binits{R.A.}},
\bauthor{\bsnm{{Tarbell}}, \binits{T.D.}},
\bauthor{\bsnm{{Wuelser}}, \binits{J.-P.}},
\bauthor{\bsnm{{Wolfson}}, \binits{C.J.}},
\bauthor{\bsnm{{Yanari}}, \binits{C.}},
\bauthor{\bsnm{{Bookbinder}}, \binits{J.A.}},
\bauthor{\bsnm{{Cheimets}}, \binits{P.N.}},
\bauthor{\bsnm{{Caldwell}}, \binits{D.}},
\bauthor{\bsnm{{Deluca}}, \binits{E.E.}},
\bauthor{\bsnm{{Gates}}, \binits{R.}},
\bauthor{\bsnm{{Golub}}, \binits{L.}},
\bauthor{\bsnm{{Park}}, \binits{S.}},
\bauthor{\bsnm{{Podgorski}}, \binits{W.A.}},
\bauthor{\bsnm{{Bush}}, \binits{R.I.}},
\bauthor{\bsnm{{Scherrer}}, \binits{P.H.}},
\bauthor{\bsnm{{Gummin}}, \binits{M.A.}},
\bauthor{\bsnm{{Smith}}, \binits{P.}},
\bauthor{\bsnm{{Auker}}, \binits{G.}},
\bauthor{\bsnm{{Jerram}}, \binits{P.}},
\bauthor{\bsnm{{Pool}}, \binits{P.}},
\bauthor{\bsnm{{Soufli}}, \binits{R.}},
\bauthor{\bsnm{{Windt}}, \binits{D.L.}},
\bauthor{\bsnm{{Beardsley}}, \binits{S.}},
\bauthor{\bsnm{{Clapp}}, \binits{M.}},
\bauthor{\bsnm{{Lang}}, \binits{J.}},
\bauthor{\bsnm{{Waltham}}, \binits{N.}}:
\byear{2012},
\batitle{{The Atmospheric Imaging Assembly (AIA) on the Solar Dynamics
  Observatory (SDO)}}.
\bjtitle{\solphys}
\bvolume{275},
\bfpage{17}.
\doiurl{https://doi.org/10.1007/s11207-011-9776-8}.
\adsurl{2012SoPh..275...17L}.
\end{barticle}
\endbibitem

\bibitem[\protect\citeauthoryear{{Leroy}}{1972}]{Leroy72}
\begin{barticle}
\bauthor{\bsnm{{Leroy}}, \binits{J.-L.}}:
\byear{1972},
\batitle{{Emissions 'froides' dans la couronne solaire}}.
\bjtitle{\solphys}
\bvolume{25},
\bfpage{413}.
\doiurl{https://doi.org/10.1007/BF00192338}.
\adsurl{1972SoPh...25..413L}.
\end{barticle}
\endbibitem

\bibitem[\protect\citeauthoryear{{Marsch} \textit{et~al.}}{2008}]{Marsch08}
\begin{barticle}
\bauthor{\bsnm{{Marsch}}, \binits{E.}},
\bauthor{\bsnm{{Tian}}, \binits{H.}},
\bauthor{\bsnm{{Sun}}, \binits{J.}},
\bauthor{\bsnm{{Curdt}}, \binits{W.}},
\bauthor{\bsnm{{Wiegelmann}}, \binits{T.}}:
\byear{2008},
\batitle{{Plasma Flows Guided by Strong Magnetic Fields in the Solar Corona}}.
\bjtitle{\apj}
\bvolume{685},
\bfpage{1262}.
\doiurl{https://doi.org/10.1086/591038}.
\adsurl{2008ApJ...685.1262M}.
\end{barticle}
\endbibitem

\bibitem[\protect\citeauthoryear{{Mart{\'\i}nez-G{\'o}mez}
  \textit{et~al.}}{2019}]{MartinezGomez19}
\begin{botherref}
\oauthor{\bsnm{{Mart{\'\i}nez-G{\'o}mez}}, \binits{D.}},
\oauthor{\bsnm{{Oliver}}, \binits{R.}},
\oauthor{\bsnm{{Khomenko}}, \binits{E.}},
\oauthor{\bsnm{{Collados}}, \binits{M.}}:
2019,
{Two-dimensional simulations of coronal rain dynamics. I. Model with vertical
  magnetic field and an unbounded atmosphere}.
\textit{arXiv e-prints},
arXiv:1911.06638.
\adsurl{2019arXiv191106638M}.
\end{botherref}
\endbibitem

\bibitem[\protect\citeauthoryear{{McIntosh} \textit{et~al.}}{2012}]{McIntosh12}
\begin{barticle}
\bauthor{\bsnm{{McIntosh}}, \binits{S.W.}},
\bauthor{\bsnm{{Tian}}, \binits{H.}},
\bauthor{\bsnm{{Sechler}}, \binits{M.}},
\bauthor{\bsnm{{De Pontieu}}, \binits{B.}}:
\byear{2012},
\batitle{{On the Doppler Velocity of Emission Line Profiles Formed in the
  ``Coronal Contraflow'' that Is the Chromosphere-Corona Mass Cycle}}.
\bjtitle{\apj}
\bvolume{749},
\bfpage{60}.
\doiurl{https://doi.org/10.1088/0004-637X/749/1/60}.
\adsurl{2012ApJ...749...60M}.
\end{barticle}
\endbibitem

\bibitem[\protect\citeauthoryear{{M{\"u}ller}, {Hansteen}, and
  {Peter}}{2003}]{Muller03}
\begin{barticle}
\bauthor{\bsnm{{M{\"u}ller}}, \binits{D.A.N.}},
\bauthor{\bsnm{{Hansteen}}, \binits{V.H.}},
\bauthor{\bsnm{{Peter}}, \binits{H.}}:
\byear{2003},
\batitle{{Dynamics of solar coronal loops. I. Condensation in cool loops and
  its effect on transition region lines}}.
\bjtitle{\aap}
\bvolume{411},
\bfpage{605}.
\doiurl{https://doi.org/10.1051/0004-6361:20031328}.
\adsurl{2003A\%26A...411..605M}.
\end{barticle}
\endbibitem

\bibitem[\protect\citeauthoryear{{M{\"u}ller}, {Peter}, and
  {Hansteen}}{2004}]{Muller04}
\begin{barticle}
\bauthor{\bsnm{{M{\"u}ller}}, \binits{D.A.N.}},
\bauthor{\bsnm{{Peter}}, \binits{H.}},
\bauthor{\bsnm{{Hansteen}}, \binits{V.H.}}:
\byear{2004},
\batitle{{Dynamics of solar coronal loops. II. Catastrophic cooling and
  high-speed downflows}}.
\bjtitle{\aap}
\bvolume{424},
\bfpage{289}.
\doiurl{https://doi.org/10.1051/0004-6361:20040403}.
\adsurl{2004A\%26A...424..289M}.
\end{barticle}
\endbibitem

\bibitem[\protect\citeauthoryear{{M{\"u}ller} \textit{et~al.}}{2005}]{Muller05}
\begin{barticle}
\bauthor{\bsnm{{M{\"u}ller}}, \binits{D.A.N.}},
\bauthor{\bsnm{{De Groof}}, \binits{A.}},
\bauthor{\bsnm{{Hansteen}}, \binits{V.H.}},
\bauthor{\bsnm{{Peter}}, \binits{H.}}:
\byear{2005},
\batitle{{High-speed coronal rain}}.
\bjtitle{\aap}
\bvolume{436},
\bfpage{1067}.
\doiurl{https://doi.org/10.1051/0004-6361:20042141}.
\adsurl{2005A\%26A...436.1067M}.
\end{barticle}
\endbibitem

\bibitem[\protect\citeauthoryear{{Parker}}{1953}]{Parker1953}
\begin{barticle}
\bauthor{\bsnm{{Parker}}, \binits{E.N.}}:
\byear{1953},
\batitle{{Instability of Thermal Fields.}}
\bjtitle{\apj}
\bvolume{117},
\bfpage{431}.
\doiurl{https://doi.org/10.1086/145707}.
\adsurl{1953ApJ...117..431P}.
\end{barticle}
\endbibitem

\bibitem[\protect\citeauthoryear{{Pesnell}, {Thompson}, and
  {Chamberlin}}{2012}]{Pesnell12}
\begin{barticle}
\bauthor{\bsnm{{Pesnell}}, \binits{W.D.}},
\bauthor{\bsnm{{Thompson}}, \binits{B.J.}},
\bauthor{\bsnm{{Chamberlin}}, \binits{P.C.}}:
\byear{2012},
\batitle{{The Solar Dynamics Observatory (SDO)}}.
\bjtitle{\solphys}
\bvolume{275},
\bfpage{3}.
\doiurl{https://doi.org/10.1007/s11207-011-9841-3}.
\adsurl{2012SoPh..275....3P}.
\end{barticle}
\endbibitem

\bibitem[\protect\citeauthoryear{{Reale} \textit{et~al.}}{2013}]{Reale13}
\begin{barticle}
\bauthor{\bsnm{{Reale}}, \binits{F.}},
\bauthor{\bsnm{{Orlando}}, \binits{S.}},
\bauthor{\bsnm{{Testa}}, \binits{P.}},
\bauthor{\bsnm{{Peres}}, \binits{G.}},
\bauthor{\bsnm{{Landi}}, \binits{E.}},
\bauthor{\bsnm{{Schrijver}}, \binits{C.J.}}:
\byear{2013},
\batitle{{Bright Hot Impacts by Erupted Fragments Falling Back on the Sun: A
  Template for Stellar Accretion}}.
\bjtitle{Science}
\bvolume{341},
\bfpage{251}.
\doiurl{https://doi.org/10.1126/science.1235692}.
\adsurl{2013Sci...341..251R}.
\end{barticle}
\endbibitem

\bibitem[\protect\citeauthoryear{{Schrijver}}{2001}]{Schrijver01}
\begin{barticle}
\bauthor{\bsnm{{Schrijver}}, \binits{C.J.}}:
\byear{2001},
\batitle{{Catastrophic cooling and high-speed downflow in quiescent solar
  coronal loops observed with TRACE}}.
\bjtitle{\solphys}
\bvolume{198},
\bfpage{325}.
\doiurl{https://doi.org/10.1023/A:1005211925515}.
\adsurl{2001SoPh..198..325S}.
\end{barticle}
\endbibitem

\bibitem[\protect\citeauthoryear{Shimizu \textit{et~al.}}{2019}]{Shimizu+19}
\begin{bchapter}
\bauthor{\bsnm{Shimizu}, \binits{T.}},
\bauthor{\bsnm{Imada}, \binits{S.}},
\bauthor{\bsnm{Kawate}, \binits{T.}},
\bauthor{\bsnm{Ichimoto}, \binits{K.}},
\bauthor{\bsnm{Suematsu}, \binits{Y.}},
\bauthor{\bsnm{Hara}, \binits{H.}},
\bauthor{\bsnm{Katsukawa}, \binits{Y.}},
\bauthor{\bsnm{Kubo}, \binits{M.}},
\bauthor{\bsnm{Toriumi}, \binits{S.}},
\bauthor{\bsnm{Watanabe}, \binits{T.}},
\bauthor{\bsnm{Yokoyama}, \binits{T.}},
\bauthor{\bsnm{Korendyke}, \binits{C.M.}},
\bauthor{\bsnm{Warren}, \binits{H.P.}},
\bauthor{\bsnm{Tarbell}, \binits{T.}},
\bauthor{\bsnm{Pontieu}, \binits{B.D.}},
\bauthor{\bsnm{Teriaca}, \binits{L.}},
\bauthor{\bsnm{Schühle}, \binits{U.H.}},
\bauthor{\bsnm{Solanki}, \binits{S.}},
\bauthor{\bsnm{Harra}, \binits{L.K.}},
\bauthor{\bsnm{Matthews}, \binits{S.}},
\bauthor{\bsnm{Fludra}, \binits{A.}},
\bauthor{\bsnm{Auchère}, \binits{F.}},
\bauthor{\bsnm{Andretta}, \binits{V.}},
\bauthor{\bsnm{Naletto}, \binits{G.}},
\bauthor{\bsnm{Zhukov}, \binits{A.}}:
\byear{2019},
\bctitle{{The Solar-C\_EUVST mission}}.
In: \beditor{\bsnm{Siegmund}, \binits{O.H.}} (ed.)
\bbtitle{UV, X-Ray, and Gamma-Ray Space Instrumentation for Astronomy XXI}
\bseriesno{11118},
\bpublisher{SPIE}, \blocation{???},
\bfpage{27 }.
\bcomment{International Society for Optics and Photonics}.
\doiurl{https://doi.org/10.1117/12.2528240}.
\burl{https://doi.org/10.1117/12.2528240}.
\end{bchapter}
\endbibitem

\bibitem[\protect\citeauthoryear{{Tian} \textit{et~al.}}{2014}]{Tian14}
\begin{barticle}
\bauthor{\bsnm{{Tian}}, \binits{H.}},
\bauthor{\bsnm{{DeLuca}}, \binits{E.}},
\bauthor{\bsnm{{Reeves}}, \binits{K.K.}},
\bauthor{\bsnm{{McKillop}}, \binits{S.}},
\bauthor{\bsnm{{De Pontieu}}, \binits{B.}},
\bauthor{\bsnm{{Mart{\'{\i}}nez-Sykora}}, \binits{J.}},
\bauthor{\bsnm{{Carlsson}}, \binits{M.}},
\bauthor{\bsnm{{Hansteen}}, \binits{V.}},
\bauthor{\bsnm{{Kleint}}, \binits{L.}},
\bauthor{\bsnm{{Cheung}}, \binits{M.}},
\bauthor{\bsnm{{Golub}}, \binits{L.}},
\bauthor{\bsnm{{Saar}}, \binits{S.}},
\bauthor{\bsnm{{Testa}}, \binits{P.}},
\bauthor{\bsnm{{Weber}}, \binits{M.}},
\bauthor{\bsnm{{Lemen}}, \binits{J.}},
\bauthor{\bsnm{{Title}}, \binits{A.}},
\bauthor{\bsnm{{Boerner}}, \binits{P.}},
\bauthor{\bsnm{{Hurlburt}}, \binits{N.}},
\bauthor{\bsnm{{Tarbell}}, \binits{T.D.}},
\bauthor{\bsnm{{Wuelser}}, \binits{J.P.}},
\bauthor{\bsnm{{Kankelborg}}, \binits{C.}},
\bauthor{\bsnm{{Jaeggli}}, \binits{S.}},
\bauthor{\bsnm{{McIntosh}}, \binits{S.W.}}:
\byear{2014},
\batitle{{High-resolution Observations of the Shock Wave Behavior for Sunspot
  Oscillations with the Interface Region Imaging Spectrograph}}.
\bjtitle{\apj}
\bvolume{786},
\bfpage{137}.
\doiurl{https://doi.org/10.1088/0004-637X/786/2/137}.
\adsurl{2014ApJ...786..137T}.
\end{barticle}
\endbibitem

\bibitem[\protect\citeauthoryear{{Xia} \textit{et~al.}}{2011}]{Xia11}
\begin{barticle}
\bauthor{\bsnm{{Xia}}, \binits{C.}},
\bauthor{\bsnm{{Chen}}, \binits{P.F.}},
\bauthor{\bsnm{{Keppens}}, \binits{R.}},
\bauthor{\bsnm{{van Marle}}, \binits{A.J.}}:
\byear{2011},
\batitle{{Formation of Solar Filaments by Steady and Nonsteady Chromospheric
  Heating}}.
\bjtitle{\apj}
\bvolume{737},
\bfpage{27}.
\doiurl{https://doi.org/10.1088/0004-637X/737/1/27}.
\adsurl{2011ApJ...737...27X}.
\end{barticle}
\endbibitem

\end{thebibliography}

\end{article} 
\end{document}